# Superelasticity and Cryogenic Linear Shape Memory Effects of CaFe$_2$As$_2$

**Authors:** John T. Sypek[1], Hang Yu[2], Keith J. Dusoe[1], Gil Drachuck[3], Hetal Patel[1], Amanda M. Giroux[1], Alan I. Goldman[3], Andreas Kreyssig[3], Paul C. Canfield[3], Sergey L. Bud'ko[3], Christopher R. Weinberger[2,4], Seok-Woo Lee[1,*]

**Affiliation:**

1. *Department of Materials Science and Engineering & Institute of Materials Science, University of Connecticut, 97 North Eagleville Road, Unit 3136, Storrs CT 06269-3136, USA*

2. *Department of Mechanical Engineering and Mechanics, Drexel University, 3141 Chestnut Street, Philadelphia, PA 19104, USA*

3. *Ames Laboratory & Department of Physics and Astronomy, Iowa State University, Ames IA 50011, USA*

4. *Department of Mechanical Engineering, Colorado State University, Fort Collins CO 80523, USA*

**Corresponding Author:**

Seok-Woo Lee

- **Address)** Department of Materials Science and Engineering & Institute of Materials Science, 97 North Eagleville Road, Unit 3136 Storrs, CT 06269-3136, USA
- **Email)** seok-woo.lee@uconn.edu
- **Telephone)** +1 860-486-8028

**Abstract:**

Shape memory materials have the ability to recover their original shape after a significant amount of deformation when they are subjected to certain stimuli, for instance, heat or magnetic fields. However, their performance is often limited by the energetics and geometry of the martensitic-austenitic phase transformation. Here, we report a unique shape memory behavior in $CaFe_2As_2$, which exhibits superelasticity with over 13% recoverable strain, over 3 GPa yield strength, repeatable stress-strain response even at the micrometer scale, and cryogenic linear shape memory effects near 50 K. These properties are acheived through a reversible uni-axial phase transformation mechanism, the tetragonal/orthorhombic-to-collapsed-tetragonal phase transformation. Our results offer the possibility of developing cryogenic linear technologies with a high precision and high actuation power per-unit-volume for deep space exploration, and more broadly, suggest a mechanistic path to a class of shape memory materials, $ThCr_2Si_2$-structured intermetallic compounds.

**Introduction**

Materials, when subjected to external loads, deform either elastically or plastically. If the deformation is elastic, the material can easily recover its shape when the external stimuli is removed since shape change is governed by the reversible stretching of atomic bonds, which simply relax with the removal of the load[1]. However, it is often the case that material shape cannot be recovered easily, as the deformation is plastic and usually involves the rearrangement of bonds in addition to simple bond stretching[2]. There is one unique case in which the shape of a material can be significantly changed and a large amount of deformation can remain, but can be restored through a reversible phase transformation that is induced through additional stimuli such as a change in temperature and the application of magnetic fields[3-5]. This type of material is called a shape memory material (SMM). Because the reversible phase transformation usually brings about a large structural recovery, it is often possible to obtain high recoverable strains, an effect that is called superelasticity (or pseudo-elasticity)[4]. Superelastic performance of SMMs is usually determined by total work (spring-back) release per-unit-volume, which corresponds to the maximum possible work done by SMMs. This maximum work release can be calculated by measuring the area below stress-strain curve between zero strain and the elastic limit. The actuation performance of SMMs is usually determined by the actuation work per-unit-volume, which is the area below a stress-strain curve within phase transformation strain. Note that the total work release is different from the actuation work because the total work release considers the total recoverable strain, but the actuation work considers only the phase transformation strain. Most crystalline shape memory metallic alloys exhibit superelasticity through a martensitic-austenitic phase transformation, which is a shear process and can produce, in general, recoverable strains up to 10 %[6-8] with only a few exceptions[9,10], but yield strengths are typically much lower than 1 GPa. Of particular note, however, single crystalline shape memory ceramic micropillars[11] exhibit a maximum elastic limit of ~10 % and a maximum yield strength of ~2.5 GPa, both of which lead to an high work release per-unit-volume. Also, their high actuation stress (1~2 GPa) and high actuation strain (5~10%) produces ultrahigh actuation work per-unit-volume (~50 MJ/m$^3$)[11]. In terms of work release and actuation work per-unit-volume, shape memory ceramic micropillars are currently regarded as the state-of-the-art SMMs[11]. This leads us to a fundamental materials science research that asks us to search for any other crystalline material systems to exhibit superelastic and actuation performances comparable or even better than shape memory ceramic micropillars. Given that superelasticity and shape memory effects are primarily the result of the atomic bond strength and the geometric relationship between martensite and austenite phases, new materials that potentially have strong atomic bonds and new phase change mechanisms must be found in order to exhibit excellent superelastic and actuation performances. Therefore, we turn our attention to novel intermetallic compounds with competing electronic, magnetic and structural transitions that are currently studied at the frontiers of material physics for properties such as superconductivity.

The ternary, intermetallic compound, $CaFe_2As_2$, which has been extensively studied as an Fe-based superconductor, has recently been found to exhibit remarkable pressure sensitivity of its crystal structure as well as its electronic/magnetic properties[12]. As shown in Fig. 1(a), this material undergoes phase transformation from the tetragonal to the collapsed tetragonal (cT) phase at room temperature for applied hydrostatic pressures just under 2 GPa and from an orthorhombic, antiferromagnetic phase to the cT phase below 50 K for pressures below 0.5 GPa. X-ray diffraction measurements have demonstrated that the abrupt reduction of lattice parameter along c-axis during transformation to the cT phase correspond to an axial strain of ~10%. $CaFe_2As_2$ has been shown to be exceptionally pressure and stress sensitive; post crystal growth, annealing and quenching experiments on $CaFe_2As_2$ and its related structures (e.g. $Ca(Fe_{1-x}Co_x)_2As_2$)[13, 14] have revealed that it is possible to tune these phase transformations by controlling the internal strain associated with compositional fluctuation or nano-precipitates, allowing the cT transition to even manifest at ambient pressures. Since the phase transformation strain is roughly 10%, it is possible to achieve total recoverable strains that exceed 10% if the material is elastically compliant enough to accommodate sufficient elastic strain as well as mechanically hard enough to suppress plasticity. Intermetallic compounds usually exhibit a high strength due to their strong directional bonding. Thus, it would be possible to obtain large recoverable strains (the sum of elastic strains and phase transformation strain) and high yield strengths simultaneously, which would lead to excellent work release per-unit-volume. Also, the large phase transformation strain and high phase transformation stress due to a strong atomic bond could lead to ultrahigh actuation work per-unit-volume. This intriguing possibility motivated us to study mechanical properties of $CaFe_2As_2$ single crystals. Due to the compound's use as a basis for high temperature superconductors, its superconducting and magnetic properties have been extensively investigated[15, 16], but $CaFe_2As_2$ has never been considered as a structural material before. Here, we report its strong potential as a structural material in terms of the relation of the cT transition to superelasticity and cryogenic shape memory behavor.

## Results

**Uni-axial compression on [0 0 1] CaFe$_2$As$_2$ micropillars**

To test the hypothesis that these materials exhibit superior superelastic properties and even can be used as shape memory compounds, we investigated the mechanical response of a solution-grown, single crystal of CaFe$_2$As$_2$ using an in-situ microcompression test in a scanning electron microscope (SEM) and in-situ cryogenic neutron scattering test under pressure. We grew, out of a Sn rich quaternary melt, a millimeter-sized plate-like single crystal that possesses mirror-like (0 0 1) and {1 0 3} facets (Fig. 1(b)). Both hydrostatic pressure and thermal expansion studies suggest that the phase transformation occurs mainly due to a structural collapse along c-axis of the tetragonal (or orthorhombic) lattice[17]. Thus, uni-axial compression tests along c-axis would be the most straightforward way to characterize either superelastic, actuation, or potential shape memory properties. Due to the plate-like geometry and small dimensions of solution-grown single crystals, it is difficult to perform conventional, bulk-scale, uni-axial mechanical tests. In addition, the irregular shapes of these as-grown crystals would result in a non-uniform stress state during loading, which could lead to un-intended plastic deformation or fracture during mechanical testing. To mitigate these issues with conventional testing, we utilize the recently developed microcompression technique that uses a focused-ion beam to create cylindrical micropillars and a flat-end nanoindenter to compress the single crystal precisely along its c-axis[18-20]. Micropillars with 2 μm diameters were fabricated on the (0 0 1) surface of the single crystal, so that our micropillars are aligned along the c-axis of tetragonal lattice (Fig. 1(c)). Then, uni-axial compression tests were performed in SEM.

First, we compressed a single micropillar up to ~11.3% strain three times, and the stress-strain curves demonstrate recoverable and repeatable responses for each cycle (Fig. 2(a)). The SEM images before and after three cycles of compression revealed no difference in height (the inset of Fig. 2(a)). Thus, this large amount of deformation is fully recovered. Also, the repeatable nature of the stress-strain response implies that there is no residual damage or evidence of a stochastic transformation. This behavior is similar with that of NiTiHf shape memory alloys that exhibit an excellent repeatability with a negligible amount of fatigue damage[21]. We also performed cyclic tests with ~10% strain and 100 cycles, and confirmed that this large deformation is fully recoverable and repeatable even after 100 cycles (See Supplementary Fig. 1 and Supplementary Note 1). Note that the shape of micropillar in the inset of Fig. 2(a) is obscured due to the ring around a micropillar. At the final stage of focused-ion beam milling, we use a thin co-centric circular pattern that, when is sometimes insufficiently wide, leaves a ring of material behind. Because the focused-ion beam is highly uniform over the pattern, however, the shape of micropillar is nearly symmetric, and the ring is not attached to a micropillar. We used the height of the right side of the micropillar, which is clearly visible, and also further confirmed that the stress-strain curves in Fig. 2(a) overlaps the stress-strain curve in Fig. 2(c), which was measured from the symmetric micropillar in Fig. 1(c) (before compression) and the inset of Fig. 2(c) (after compression). Also, we always carefully inspect the measured displacement from in-situ deformation movies. Thus, all stress-strain data in this manuscript were correctly measured (Also See Supplementary Fig. 2 and Supplementary

Note 2). Additional stress-strain curves are also available in Supplementary Fig. 3, which confirm that our method produces reasonably consistent superelasticity results.

The stress-strain curves exhibit three distinct deformation behaviors (stage I, II, and III) (Fig. 2(a)). Note that the existence of non-linear behavior (stage II) resembles that of typical SMMs, which corresponds to a phase transformation[7, 11]. Another feature that is notable in the stress-strain curves is the relative narrowness of the hysteresis loop, which is also quantitatively different from standard SMMs that exhibit substantially larger hysteresis loops at room temperature. Thus, $CaFe_2As_2$ would be able to release the large amount of work without significant damping effects. Most crystalline SMMs exhibit a non-linear response through the martensitic-austenitic phase transformation. Because this shear transformation almost always produces a localized shear strain on the micrometer scale, a kinked shape is easily observed during compression on a single crystalline micropillar of conventional shape memory materials[11]. For $CaFe_2As_2$, however, we confirmed via our in-situ mechanical test of various micropillars observed along different viewpoints and at a high magnification (up to 7000 X) that $CaFe_2As_2$ always shows only a height reduction along the c-axis (loading direction) without any lateral motion of the pillar, the formation of slip steps, or strain bursts. These results imply that phase transformation is not related to a shear process at all (Fig. 2(a)) (See also Supplementary Movie 1) but rather a simple reduction in lattice constant along c-axis that is expected for this compound undergoing a cT phase transition. Thus, superelasticity in $CaFe_2As_2$ is exhibited by non-martensitic-austenitic phase transformation. In our experiment, shear-like phenomena are only observed in conjunction with plastic yielding (See the formation of shear slip steps in the inset of Fig. 2(c)). Also, we confirmed that as long as the micropillar diameter is large enough to neglect the FIB effect damage layer (typically, pillar diameter > 1 μm), $CaFe_2As_2$ micropillars exhibit no size dependence in stress-strain curve. This would be because the unit length scale of the phase transition is only the dimension of the unit cell, which is far smaller than the micropillar diameter.

**Density functional theory calculations on superelasticity**

In order to provide additional insight into the unusual form of superelasticity in $CaFe_2As_2$, we performed the density functional theory (DFT) simulations of the compression process. Since DFT calculations are done at 0 K, the stable phase of $CaFe_2As_2$ predicted is the anti-ferromagnetic orthorhombic phase, which is different from the paramagnetic tetragonal phase of our room temperature experiments[22]. As revealed previous work on the phase transformation in $CaFe_2As_2$, the formation of As-As bonds has been identified as the primary process to responsible for the structural collapse regardless of initial structure (tetragonal or orthorhombic)[22]. Thus, it is still possible to capture the essential physics of how the formation of As-As bonds affects stress-strain curves and the stability of the cT phase at a low temperature. This information will be critical for understanding of unusual shape memory effects at an ultracold temperature later. Our DFT simulations of a single unit cell, Supplementary Fig. 4 show that initially the $CaFe_2As_2$ crystal undergoes elastic deformation in stage I and collapses into the cT phase in stage II, as a consequence of new bonding formation between As-As layers and the loss of all magnetism[12]. An electron density distribution map, (Fig. 2(b)), shows the electron density overlap of As-As bonding under uni-axial compression. At this moment, bonding-antibonding splitting of As $4p_z$ orbitals is increased and this is an important indication that a bond formation has taken place[23]. The stress-strain curve of unit cell (Supplementary Fig. 4) shows the sudden decrease in stress, which indicates the sudden change (phase transformation) in structure and properties. The strong attraction force between As-As layers collapses the lattice structure. However, the simulations of a single unit cell miss the gradual phase transformation expected in the real material. To simulate this realistic case, we use a composite model that minimizes the total enthalpy under load control, similar to the experiments, to determine the phase fractions of the orthorhombic and cT phases via minimization (See Supplementary Figs. 5, 6 ,7 and Supplementary Note 3). The resulting stress-strain curve is shown in Fig. 2(c). DFT simulations, in conjunction with a simple analytical model, demonstrate that in the course of compression, the partial phase transformation is energetically more favorable than the full/abrupt transformation of the entire volume of specimen. Thus, the volume fraction of cT phase gradually increases in the course of deformation, leading to continuous stress-strain curve in Fig. 2(c). The unit cell still undergoes the abrupt structural collapse locally, but the macroscopic response of micropillar exhibits the gradual and continuous deformation. Note that the cT transition does not have an invariant plane, which is undistorted and unrotated plane. Due to small in-plane strain on a (0 0 1) plane during the cT transition as seen in Fig. 2(b), however, a (0 0 1) plane would be a preferable interface between tetragonal (or orthorhombic) and collapsed tetragonal phases. The details of microstructural evolution could be confirmed by by performing in-situ electron back-scattered diffraction (EBSD) or in-situ transmission electron microscope (TEM) micropillar compression test.

Despite the large number of assumptions in our model, it clearly captures all phases of deformation including the smooth stage II transition from orthorhombic to cT phase.

Indeed, the formation of As-As bonding (phase transformation) is the major process that induces the non-linearity of stress-strain curve. Note that superelasticity mechanism in $CaFe_2As_2$ is unique because it is not exhibited by the martensite-austenite phase transformation of conventional shape memory alloys and ceramics. From these single unit cell and composite simulations, it is clear that the large values of recoverable strain observed in $CaFe_2As_2$ occur by the combination of linear elastic strain, hyperelastic strain and phase transformation strain.

**Superelasticity and actuation performances**

It is clear from our experiments that we have yet to observe the maximum recoverable strain, the yield strength, the maximum work release per unit volume, and the actuation work per-unit-volume of this material. To determine these values in compression, we performed an additional micropillar compression test, increasing stress until we see either plastic deformation or fracture, and determined that the elastic limit is ~13.4 % and the yield stress is ~3.7 GPa, both of which is even higher than those of single crystalline shape memory ceramic micropillars, which have been known to possess the current state-of-the-art superelastic work and actuation work per-unit-volume (Figs. 2(c) and 3). Returning to the composite material model we see that at the experimentally determined yield stress, the simulation suggests ~13% recoverable strain, which agrees well with our experimental results. Intermetallic compounds typically possess strong directional atomic bonds, which explain why the strength is much higher than shape memory alloys or even shape memory ceramics. In fact, the strong directional bonds help suppress plastic deformation in CaFe$_2$As$_2$, increasing the recoverable strain from roughly ~7% associated with the phase change to approximately 13.4% which includes elastic strain.

The combination of the high recoverable strains (13.4%) and high yield strengths (3.7 GPa) are even higher than those of shape memory ceramic micropillars[11]. Also, low hysteresis area allows CaFe$_2$As$_2$ to absorb and release of ultrahigh work per-unit-volume with minimal dissipation at room temperature. The stress-strain curves in Fig. 2(a) suggest that the energy loss by damping is about 10% based on the ratio of the areas under the loading-to-unloading curves. Assuming that this percent loss is similar up to elastic limit, we can estimate the maximum possible work asorption/release per–unit-volume by using 90% of the area under the stress-strain curves beween zero strain and the elastic limit in Fig. 2(c): this corresponds to 1.76×10$^8$ J/m$^3$ (experiment) and 1.57×10$^8$ J/m$^3$ (DFT) for the absorption of, and 1.58×10$^8$ J/m$^3$ (experiment) and 1.40×10$^8$ J/m$^3$ (DFT) for the release of work per-unit-volume (for 10% of damping effect). These results imply that CaFe$_2$As$_2$ is able to release the superelastic work about 10~1000 times larger than most engineering materials (e.g. Stainless steel[24]: 10$^5$ J/m$^3$, Zr-based bulk metallic glass[25]: 2×10$^7$ J/m$^3$, shape memory ceramic micropillars[11]: 4×10$^7$ J/m$^3$: Shape memory alloys[11]: 10$^5$ J/m$^3$). Some shape memory metallic alloys could show the larger recoverable strain. Due to their low yield strength and high damping effect, however, their work release per-unit-volume (~10$^5$ J/m$^3$) is much lower than CaFe$_2$As$_2$ (1.76×10$^8$ J/m$^3$). Furthermore, as noted before, the actuation capability is determined by the maximum possible actuation work per-unit-volume which is the area below a stress-strain curve within phase transformation strain (the grey trapezoidal area in Fig. 2(c)), which corresponds to ~5.15×10$^7$ J/m$^3$. This can be estimated by multiplying the actuation strain by the average actuation stress ($\sigma_{PT}^{ave}$), which is the average of the onset ($\sigma_{PT}^{onset}$) and offset ($\sigma_{PT}^{offset}$) phase transformation stresses (Fig. 2(c)). Based on this scheme, we can compare the actuation capability of CaFe$_2$As$_2$ with other actuation systems by locating our four micropillar results and one DFT result with the actuation stress and the average actuation stress (See Supplementary Fig. 8, Supplementary Table 1, and Supplementary Note 4). Fig. 3 shows a comparison of actuation capability, by plotting actuation stress versus actuation strain, of CaFe$_2$As$_2$ compared with other materials[11, 26, 27]. Note that materials with higher actuation work per-unit-volume appear towards the top and right of the plot in Fig. 3. Conventional shape memory alloys, such as bulk Ni-Ti alloy, exhibit much lower actuation work per-unit-volume, as also evidenced by their low actuation stress (see the grey colored stress-strain curve in Fig. 2(a)). Norfleet *et al.* investigated shape memory properties of Ni-Ti micropillars[28]. Size effects can improve the phase transformation stress of Ni-Ti alloys, up to 800 MPa. Also, the transformation strain of Ni-Ti micropillar is around 0.06. Then, actuation work per-unit-volume of Ni-Ti micropillar (800 MPa × 0.06 = 4.8×10$^7$ J/m$^3$) is similar with that of CaFe$_2$As$_2$ (~5.15×10$^7$ J/m$^3$). Fig. 3 clearly shows that CaFe$_2$As$_2$ shows the comparable actuation work per-unit-volume with shape memory ceramic micropillars (the current state-of-the-art) or the high-end of shape memory alloy. Thus, CaFe$_2$As$_2$ exhibit excellent actuation work per-unit-volume, compared to shape memory alloys.

**Cryogenic linear shape memory behavior**

The narrow hysteresis loop of $CaFe_2As_2$ would be associated with the different stress required to form and break the As-As bonds. This leads to the existence of hysteresis in stress-strain curve of $CaFe_2As_2$. Our DFT simulation results (Supplementary Fig. 4) and Goldman *et al.*'s neutron scattering results (Fig. 4(a)) show that $CaFe_2As_2$ exhibits the maximum width of hysteresis at 0 Kelvin[22]. Without thermal vibrations, the As-As bonds are not broken even after the complete removal of compression stress. The neutron scattering experiment (Fig. 4(a)) show that a hydrostatic tensile load of 0.25 GPa (or -0.25 GPa pressure) is required to destabilize cT phase at 0 K[22]. As the temperature increases, thermal vibrations help the destabilization of the As-As bonds under some amount of compression stress (or pressure), leading to the narrower hysteresis. Fig. 4(a) shows clearly that the width of hysteresis area becomes narrower as the temperature increases. Thus, increasing a temperature would be an effective way to make the hysteresis loop narrower and to maximize the release of mechanical work. Vice versa, it is possible to expect the cryogenic shape memory effects due to the metastability of cT phase at 0 K.

The DFT simulation of a single unit cell predicts that the cT phase is metastable after loading-unloading cycle at 0 K (Supplementary Fig. 4). In other words, cT phase is not transformed into orthorhombic phase even though the applied stress is completely removed at 0 K. However, our experimental data shows the full strain recovery at room temperature, suggesting that the cT phase is unstable at room temperature but may well be metastable near 0 K at atmospheric pressure. Metastability of cT phase is a pre-requisite to exhibit shape memory behavior. Thus, this work includes important evidence that, taken together with our current superelasticity results, suggests the existence of a cryogenic linear shape memory effect. Based on this evidence, it is possible to suggest, two different deformation routes, path 1-2-3-4 and path 1'-2', to demonstrate the cryogenic linear shape memory effects and thermal actuation, respectively (Fig. 4). In order to prove the existence of cryogenic shape memory effects, we conducted in-situ cryogenic neutron scattering measurement on a single crystal of $CaFe_2As_2$ using a He-gas pressure cell in a displex cryogenic refrigerator[22]. In this experiment, it is possible to control both temperature and pressure and to check the occurrence of phase transformation at the same time (Fig. 4(a)). First, the temperature was decreased down to 50 K, and neutron scattering data clearly shows that the phase becomes orthorhombic (path 1 and Fig. 4(b)). Then, the pressure was increased until cT transition occurs (path 2 and Fig. 4(b), 4(c)), and we confirmed that cT transition occurs between 350 and 400 MPa. A prerequisite of the shape memory effect is metastability of the deformed state. To achieve this metastability of cT phase, we decreased the pressure down to 300 MPa (path 3), and neutron scattering data confirms that the transition to the T phase does not occur even below the critical pressure for the forward transition (350~400 MPa) (Fig. 4(d)). The transition to the O phase (at 50 K) only occurs when the pressure is decreased below 50 MPa (Fig. 4(e)). This clearly confirms that cT phase is metastable after path 3 and satisfies the pre-requisite of shape memory effect, which is metastable deformed state. After preparing the sample in the cT phase, within the hysteresis region, the temperature was increased from 50K to 100K to destabilize the cT phase, and the transition to the orthorhombic phase occurs (path 4 in Fig. 4(a) and Fig. 4(f)). This is truly shape memory effect that exhibits the shape recovery by making the metastable deformed state unstable by applying the stimulus, heat (Fig 4(h)). The transformation strain along a- and b-axes are much smaller than the transformation strain along c-axis. Thus, the cT transition is nearly linear volume transformation (the volume changes only by the lattice expansion/contraction along c-axis). Thus, cryogenic linear shape memory effects do exist in $CaFe_2As_2$. Based on the phase diagram (Fig. 4(a)), the cryogenic linear shape memory effects can be shown at the pressure between 0 and 400 MPa and at the temperature between 0 K and 110 K. Thus, theoretically, it is possible to achieve shape memory effects near 0K, and this property would be useful to develop a cryogenic linear actuator working even in deep space (3 K). In addition, path 1' and 2' also shows the thermal actuation at a temperature higher than 100 K by the phase transformation between tetragonal and cT phase. Neutron scattering data clearly shows that this transformation occurs and thermal actuation is available (Fig. 4(g)) as shown in schematic diagram (Fig. 4(i)).

Whereas this shape memory behavior is based on experimental observations during hydrostatic loading, our uniaxial results suggest that it is a real possibility in uniaxial compression as well which calls for further exploration. Also note that the phase transformation in $CaFe_2As_2$ accumulates almost no residual damage as seen in the repeatable stress-strain curves in Fig. 2(a), thus ensuring dimensional stability. The phase transformation in $CaFe_2As_2$ occurs simply by the formation and breakage of As-As bonding. Thus, severe stress concentrations, the build-up of dislocation structures or residual stresses, which are usually observed in the martensitic-austenitic phase transition, are not expected to occur. Additionally, we conducted 100 cycles of forward-backward transition, and the post-mortem SEM confirmed that there is no residual deformation after 100 cycles (Supplementary Fig. 1). Due to the limitation of small-scale mechanical testing, such as the thermal drift issue, we were not able to perform extensive cyclic test with milliions of cycles as usually done for conventional shape memory alloys. However, we believe CaFe$_2$As$_2$ would exhibit superior fatigue resistance due to the simplicity of breaking and making bond mechanism, which is not too different from bond stretching of normal elastic deformation. This implies that it is possible to develop a cryogenic actuator system with high actuation power as well as negligible degradation of actuation performance. Also, the repeatability of the stress-strain response even at the micrometer scale ensures ultra-high precision for repeatable actuation motion. Note that the small-dimensions and single crystalline nature of our samples in this study is certainly beneficial in achieving the higher performance observed here, similar to that seen in shape memory ceramics[11]. Also, it is important in this material system to use either hydrostatic pressure or uniform uni-axial stress along c-axis to avoid plastic deformation and fracture during the actuation process.

**Discussion**

The observation of superelastic and possible shape memory behavior in CaFe$_2$As$_2$ opens a number of possibilities because the phase transformation can be tuned by adjusting the chemical composition and by controlling the internal strain. For instance, the substitution of cobalt for iron coupled with annealing and quenching can induce the significant shift of boundaries in phase diagram in Fig. 1(a). Thus, it is possible to tune the phase diagram to fit specific needs by engineering the chemistry as well as the microstructure of this class of materials. Furthermore, and perhaps of even greater importance, the cT phase transition found in CaFe$_2$As$_2$ is only one manifestation of a wider class of such transitions found in over 400 different ThCr$_2$Si$_2$-type intermetallic compounds. The novelty of this bond making / breaking phase transition was appreciated as early as 1985 when Hoffmann and Zheng surveyed a number of potential systems that could manifest this type of transition[29]. Based on our results on $CaFe_2As_2$, we fully anticipate that other members of the $ThCr_2Si_2$ structure class will manifest similar superelasticity and potential as SMM. Any $ThCr_2Si_2$-type intermetallic compounds that can undergo the cT transition under compression can exhibit superelasticity in the same way as long as they are not plastically yielded or fracture before the phase transformation occurs. Some intermetallic compounds exist as a collapsed tetragonal state under stress-free condition[28]. In this case, superelasticity can be shown under tension by breaking Si-Si type bonds, but the same principle from our results works on this case, too. It is certainly beneficial to search for a new superelastic and shape memory materials that do not contain a toxic arsenic (As), which might make a poor impact to industry. Indeed, our recent experiments and DFT simulations confirmed that $LaRu_2P_2$, one of $ThCr_2Si_2$-type intermetallic compound, also exhibits the mechanical responses similar with those of $CaFe_2As_2$ (See Supplementary Fig. 9 and Supplementary Note 5).

In sum, our results suggest the strong potential of $CaFe_2As_2$ as a structural material in terms of high elastic energy storage, high actuation power, low fatigue damages, and cryogenic shape memory behavior. Forming and breaking As-As type bond process of cT transition is an entirely different from a conventional martensitic-austenitic transformation process. Our results offer the possibility of developing cryogenic linear technologies with a high precision and high actuation power per-unit-volume for deep space exploration, and more broadly, suggest a mechanistic path to a whole new class of shape memory materials, $ThCr_2Si_2$-structured intermetallic compounds.

**Materials and Method**

**Single crystal growth.** Single crystals of $CaFe_2As_2$ were grown out from a Sn flux, using conventional high-temperature solution method[30]. Elements Ca, Fe, As and Sn were combined together in the ratio of Ca : Fe : As : Sn = 2 : 3.5 : 4 : 96. Single crystals were grown by slowly cooling the Ca-Fe-As-Sn melt from 1180 °C to 600 °C at 5 °C/h, and then decanting off the excess liquid flux. Detailed description of the crystal growth can be found elsewhere[31]. The obtained crystals have typical size up to 5x5x1 $mm^3$.

**In-situ nanomechanical measurement.** Micropillars were produced using an FEI Helios Nanolab 460F1 FIB machine. Gallium ion beam currents from 300 to 10 pA under an operating voltage of 30 kV were used from the initial to final thinning with concentric circle patterns. Note that the typical thickness of FIB damage layer is about 20 nm, which is much thinner than our pillar diameter (~2 μm). Thus, the effects of FIB damage layer on mechanical data should be negligible. Also, in our study, the size effect in superelasticity is not expected since the length scale of phase transformation is unit lattice size, which is much smaller than our pillar diameter (~2 μm). Thus, our micropillar volume should be the representative volume of the bulk. In-situ nanomechanical test was performed at room temperature and under ultra-high vacuum condition ($<10^{-4}$ Pa) using NanoFlip™ (Nanomechanics, inc., TN, USA), which is installed in a field-emission gun JEOL 6330F scanning electron microscope (JEOL, Japan). The nominal displacement rate of 10 nm/s, which corresponds to the engineering strain rate of 0.002 $s^{-1}$, was used for all in-situ compression tests in this study. The thermal drift was always smaller than 0.01 nm/s, which ensures the negligible error in displacement measurement.

**Density functional theory calculation.** The density functional theory simulations were carried out using the Vienna Ab-initio Simulation (VASP) package[32] using a plane wave basis. The projector augmented wave (PAW) psuedopotentials[33] were used in all calculations and the exchange-correlation energies were evaluated using the formulation of Perdew, Burke and Ernzerhoff (PBE)[34] within the generalized gradient approximation (GGA). The energy cutoff of the plane waves was set to 450 eV and a Monkhorst Pack integration scheme of $10 \times 10 \times 5$ was used for the integration of the irreducible Brillouin zone. Most of the calculations involve the deformation of a single unit cell of $CaFe_2As_2$. At low temperatures, the orthorhombic structure is the stable structure with 20 atoms in its unit cell and this structure was used in all our calculations, even for non-magnetic collapsed tetragonal case. The lattice constants of the orthorhombic phase were found to be 5.60 Å, 5.49 Å and 11.48 Å and the iron moment was found to be 1.83 $\mu_B$. The lattice constants agree well with experimental values while the iron moment is large, but these findings are in very good agreement with previous DFT studies[35, 36, 37].

**Neutron scattering measurement.** We obtained neutron scattering data by using the BT-7 spectrometer of the NIST Center for Neutron Research. The scattering data were obtained in double-axis mode with a wavelength of 2.36 Å. Higher harmonic portion of the neutron beam was eliminated by using pyrolytic graphite filters. A single crystal was wrapped in Al-foil, and then was located in a flat plate of the Al-alloy He-gas pressure cell. Then, a closed-cycle refrigerator was used to decrease the temperature of a crystal. The $(hhl)_{Tet}$ reciprocal lattice plane of the sample was exactly matched with the diffractometer scattering plane. Then, hydrostatic pressure was applied to a pressure cell. We were able to monitor and adjust the hydrostatic pressure by using a pressurizing intensifier. This system enabled us to control the pressure at fixed temperatures (above the He solidification line), or to scan the temperature at almost constant pressures. We determined the phase boundary between the orthorhombic and the collapsed tetragonal phases by changing pressure at a fixed temperature while monitoring the $(004)_{cT}$ peaks and the $(004)_{Ortho}$ diffraction peak. Also, we obtained the 'collapsed tetragonal'-'tetragonal' and 'collapsed tetragonal'-'orthorhombic' phase boundaries by measuring the intensity of the $(004)_{cT}$ diffraction peak while increasing the temperature, or the intensity of the $(004)_{Tetra}$ and $(004)_{Ortho}$ diffraction peaks while decreasing the temperature.

**Data availability**

All relevant data are available from the corresponding author on request.

**Acknowledgements**

J.T. Sypek, K.J. Dusoe and S.-W Lee acknowledge support from the UConn Start-up Funding, the UConn Research Excellence Program Funding, and the Early Career Faculty Grant from NASA's Space Technology Research Grants Program. FIB work was performed using the facilities in the UConn/FEI Center for Advanced Microscopy and Materials Analysis (CAMMA). Work by P.C. Canfield and S.L. Bud'ko was supported by the U.S. Department of Energy, Office of Basic Energy Science, Division of Materials Sciences and Engineering. Their research was performed at the Ames Laboratory. Ames Laboratory is operated for the U.S. Department of Energy by Iowa State University under Contract No. DE-AC02-07CH11358.

**Author Contributions**

J.T.S. and S.-W.L. conceived and designed the experiments and simulations, and performed in-situ micro-mechanical tests. P.C.C. and S.L.B. proposed these measurements and grew a $CaFe_2As_2$ single crystal. P.C.C., S.L.B., and G.D. grew a $LaRu_2P_2$ single crystal. H.Y. and C.R.W. performed density functional theory calculation. H.P., A.M.G., and K.J.D. performed nanoindentation and in-situ micro-mechanical test on $LaRu_2P_2$. A.I.G., A.K., P.C.C. and S.L.B. performed cryogenic neutron scattering measurement under pressure. J.T.S., P.C.C., C.R.W., and S.-W.L. wrote the paper. All authors discussed the results and commented on the manuscript. Correspondence should be addressed to S.-W.L.

**Competing interests**

The authors declare no competing financial interests.

**Fig. 1. Intermetallic compound $CaFe_2As_2$** (a) The phase diagram of $CaFe_2As_2$ in temperature-pressure space[12], (b) an optical microscope image of solution-grown single crystalline $CaFe_2As_2$; scale bar, 1 mm (c) a [0 0 1]-oriented $CaFe_2As_2$ micropillar with 2 μm in diameter made by Ga+ focused-ion beam milling; scale bar, 1 μm

**Fig. 2. Demonstration of superelasticity including large recoverable strains and high yield strengths** (a) A three stress-strain curves of same [0 0 1]-oriented $CaFe_2As_2$ micropillar; the inset shows no height change even after three cycles of 11.3 % strain compression. Remarkably, the stress-strain curves appear identical, indicating that there is no residual damage accumulation during cyclic deformation, which is typically observed in shape memory alloys or ceramics especially at small length scales[7, 11]. Stress-strain data of Ni-Ti alloy also was included for comparison. Two snapshots of in-situ compression show the significant compressive elastic deformation, which cannot be achieved in other metallic or intermetallic materials; scale bar, 1 μm. (b) Electron density distributions at ambient pressure and under c-axis compression; the formation of As-As bonding, which is indicated by arrows, induces the structural collapse and the 10% height reduction. The iso-surface level used was 0.035 and the max is 0.1 (red). (c) Experimental and computational (DFT) stress-strain curves of [0 0 1]-oriented $CaFe_2As_2$ micropillar up to the yield point. We intentionally terminated the stress-strain data of DFT simulation at the experimental yield strength because the remaining data above the yield strength is not meaningful in the current analysis. The inset shows the plastic slip in the $[30\bar{1}]/(103)$ slip system.

**Fig. 3. A comparison of CaFe$_2$As$_2$ with other structural and actuator materials**

Actuator stress and strain for various actuator materials and systems. Contours of constant specific work are indicated by dashed lines (adapted from Lai, A., *et al. Science* **27,** 1505-1508 (2013)[11], Huber, J. E., *et al. Proc. R. Soc. London Ser. A* **453,** 2185–2205 (1997)[26] and Lang, J., *et al. Adv. Mater.* **28,** 10236-10243 (2016)[27]). Both experimental and DFT data are included. All four experimental data came from stress-strain curves in Supplementary Fig. 3(c). Note that CaFe$_2$As$_2$ exhibit actuation capapbility comparable to shape memory ceramic micropillars.

**Fig. 4. Linear shape memory effect and thermal actuation in cryogenic environments**

(a) The temperature-pressure phase diagram for CaFe$_2$As$_2$[22], neutron scattering data of (b) (004)$_O$ plane of orthorhombic phase at T=50 K with increasing pressure, (c) (004)$_{cT}$ plane of collapsed tetragonal phase at T=50 K with increasing pressure, (d) (004)$_{cT}$ plane of collapsed tetragonal phase at T=50 K with decreasing pressure, (e) (004)$_O$ plane of orthorhombic phase at T=50 K and p=50 MPa, (f) (004)$_{cT}$ plane of collapsed tetragonal phase at p=300 MPa with increasing temperature, (g) (004) planes of tetragonal and collapsed tetragonal phases at p=470 MPa with increasing and decreasing temperature. Schematic diagrams of (h) one way linear shape memory effect and (i) thermal actuation.

# Supplementary Information

## Superelasticity and Cryogenic Linear Shape Memory Effects of CaFe$_2$As$_2$

**Authors:** John T. Sypek[1], Hang Yu[2], Keith J. Dusoe[1], Gil Drachuck[3], Hetal Patel[1], Amanda M. Giroux[1], Alan I. Goldman[3], Andreas Kreyssig[3], Paul C. Canfield[3], Sergey L. Bud'ko[3], Christopher R. Weinberger[2,4], Seok-Woo Lee[1,*]

**Affiliation:**

1. *Department of Materials Science and Engineering & Institute of Materials Science, University of Connecticut, 97 North Eagleville Road, Unit 3136, Storrs CT  06269-3136, USA*

2. *Department of Mechanical Engineering and Mechanics, Drexel University, 3141 Chestnut Street, Philadelphia, PA 19104, USA*

3. *Ames Laboratory & Department of Physics and Astronomy, Iowa State University, Ames IA  50011, USA*

4. *Department of Mechanical Engineering, Colorado State University, Fort Collins CO 80523, USA*

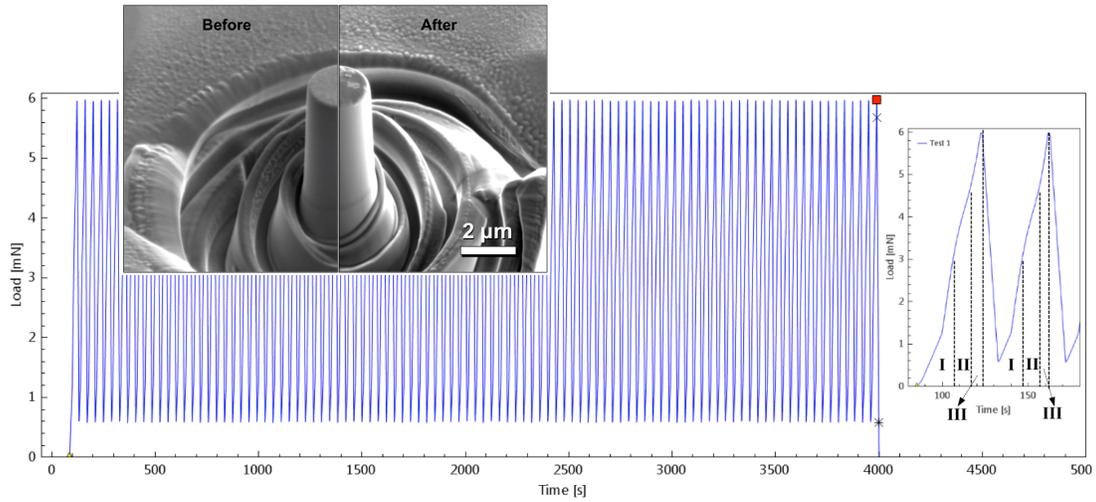

**Supplementary Figure 1: 100 cyclic compression test.** (a) The time-load curve of 100 cyclic compression test. The non-linear curve of the first two cycles in the inset clearly shows that cT phase transformation occurs every cycle. Inset is $CaFe_2As_2$ micropillar before and after 100 cycles of forward-backward transformation. This image shows that there is no residual plastic deformation after 100 cycles.

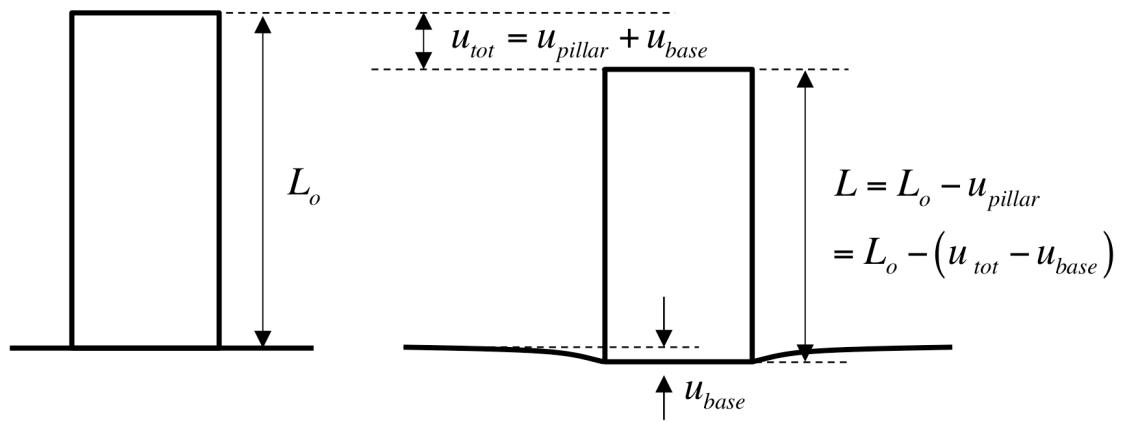

**Supplementary Figure 2: Schematic diagram of the deformed micropillar.** In order to obtain the precise strain value in the micropillar, it is important to measure the displacement of the base under the micropillar because the total measured displacement ($u_{tot}$) includes both the displacement of micropillar ($u_{pillar}$) and the base ($u_{base}$).

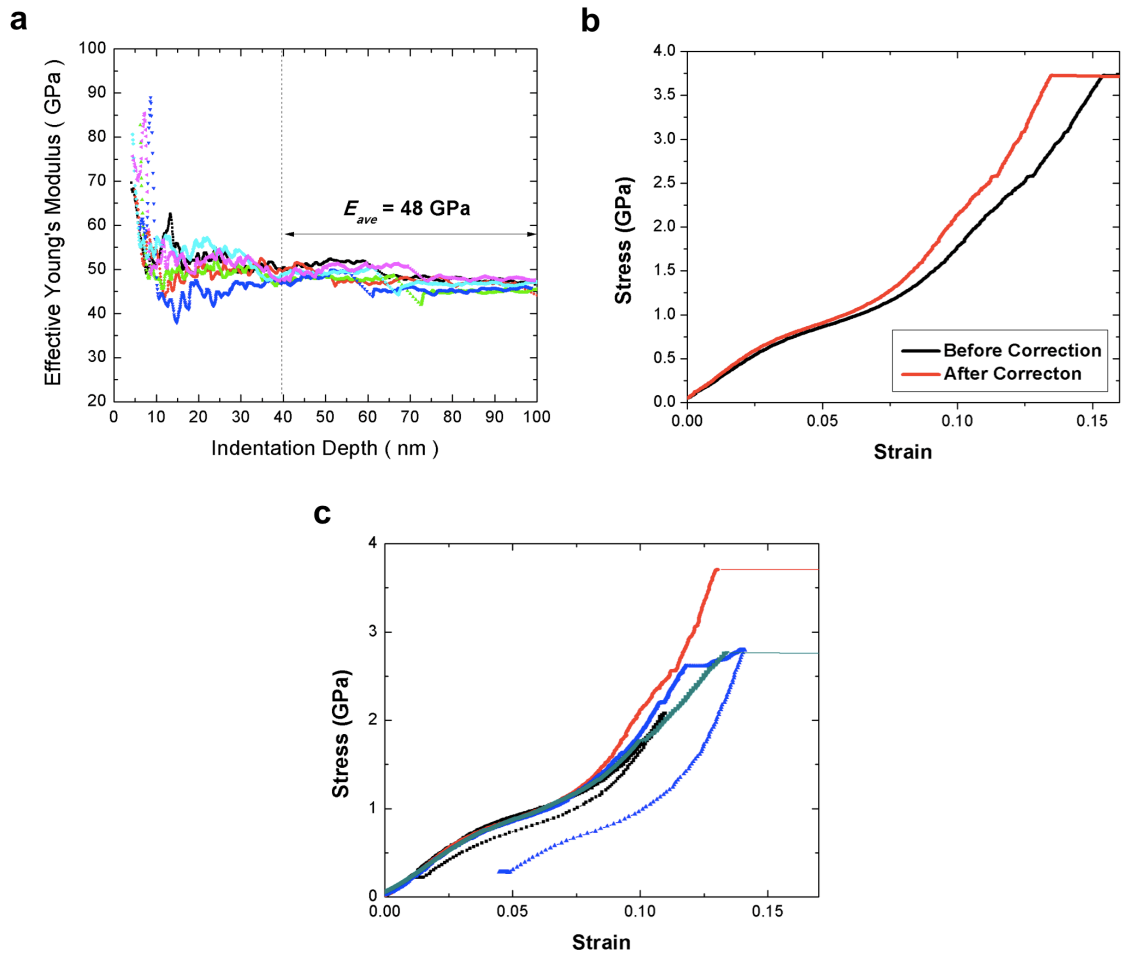

**Supplementary Figure 3: Strain correction via Sneddon's flat punch method.** (a) The measurement of the effective Young's modulus of [0 0 1] $CaFe_2As_2$, and (b) the correction of stress-strain curve by considering the displacement of the micropillar base (c) Stress-strain curves of four different samples. Note that four stress-strain curves show consistent superelasticity results.

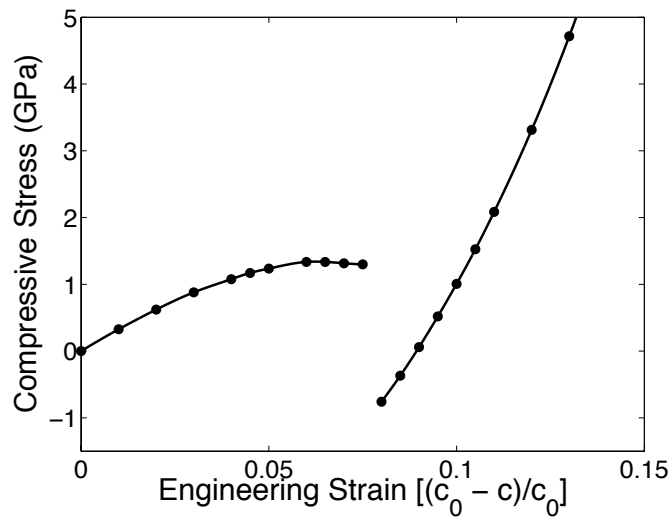

**Supplementary Figure 4: A uni-axial stress-strain curve of a single unit cell of CaFe$_2$As$_2$.** At about 8% uniaxial compression, the cell changes to the collapsed tetragonal unit cell with a reduced c-axis length. The reference value for $c_0$ is taken as the equilibrium value for c in the orthorhombic structure. The stress changes abruptly and the moment on the iron atoms vanishes.

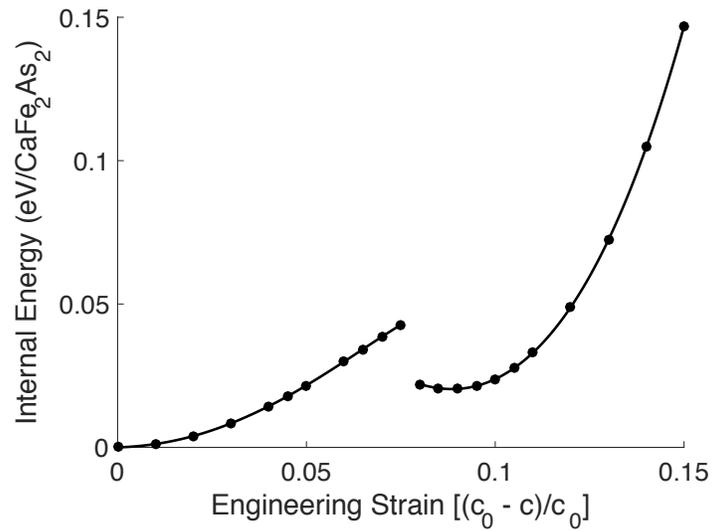

**Supplementary Figure 5: A plot of the internal energy of CaFe₂As₂ under uniaxial loading.** The energy is evaluated at each indicated strain and is plotted per formula unit of CaFe₂As₂. Note that the lateral stresses were allowed to relax during uniaxial loading to relieve the lateral stresses.

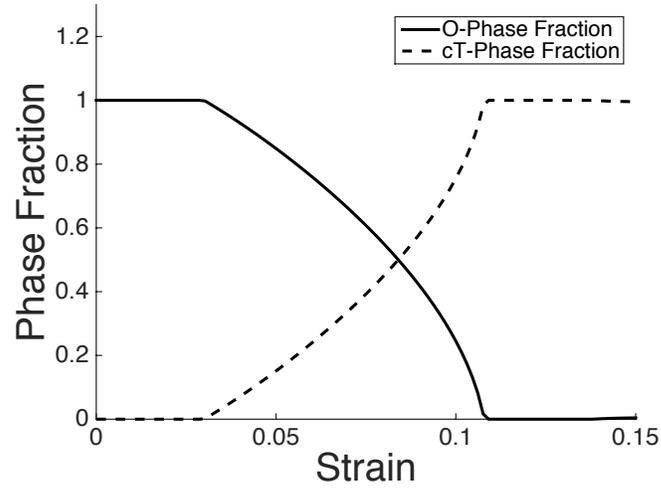

**Supplementary Figure 6: A plot of the phase fractions of the two phases as a function of the total strain.** The phase fraction evolves during the phase transition and the two phase fractions add to a total of 1.0.

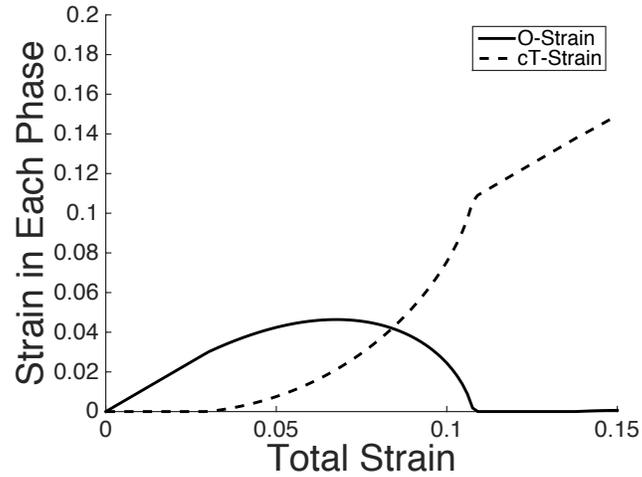

**Supplementary Figure 7: A plot of the evolution of the strain in each phase as a function of the total strain.** The engineering strain used here is defined relative to the c-axis of the Orthorhombic structure shown in Supplementary Fig. 4 and thus is not the elastic strain in the cT phase but is computationally easier to use. In this case, the strains in the two phases must add to the total strain.

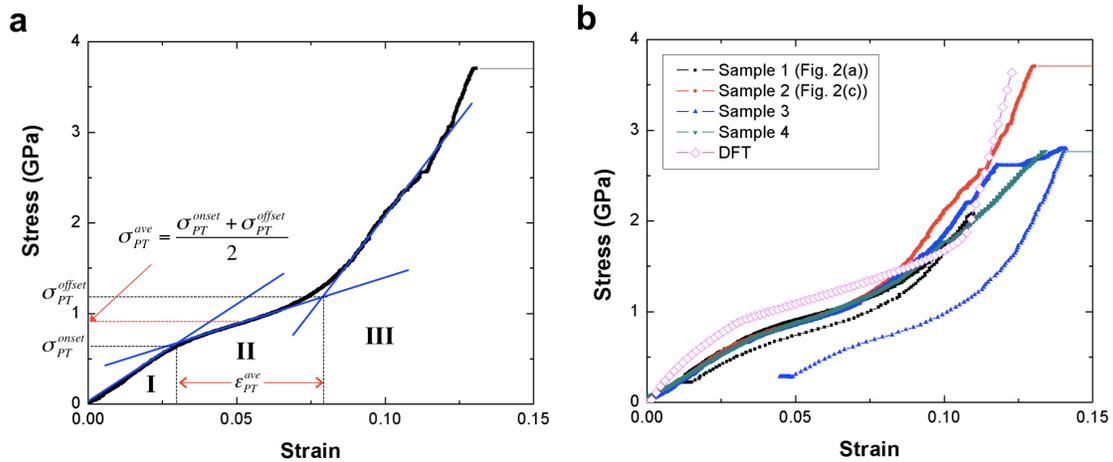

**Supplementary Figure 8: Analysis of onset and offset stress of phase transformation.** (a) Determination of $\sigma_{PT}^{onset}$, $\sigma_{PT}^{offset}$, and $\varepsilon_{PT}^{ave}$, (b) four experimental and one computational stress-strain data. Note that Sample 3 underwent the plastic deformation, and did not fully recover to the original shape. However, the elastic portion (stage I, II, III) is full recovered.

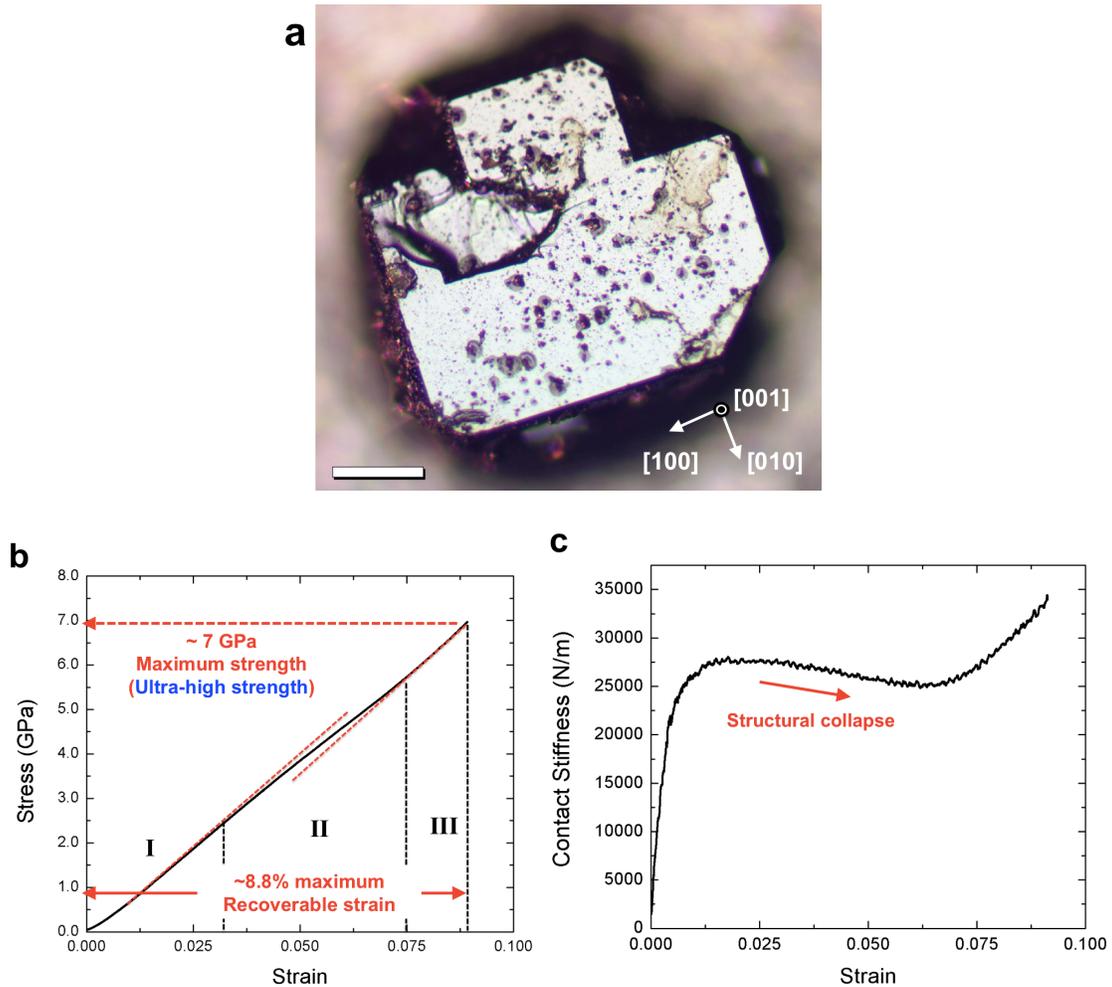

**Supplementary Figure 9: Superelastic intermetallic compound, LaRu$_2$P$_2$.** (a) Optical image of single crystalline LaRu$_2$P$_2$; scale bar, 100 μm (b) Stress-strain curve of [0 0 1] LaRu$_2$P$_2$. LaRu$_2$P$_2$ also exhibit the distinct three stages of deformation, which is the characteristic of superelastic and shape memory material (*not published*). (c) Contact stiffness as a function of strain. The decrease in contact stiffness implies that LaRu$_2$P$_2$ is collapsed during deformation.

|  | $\sigma_{PT}^{onset}$ (MPa) | $\sigma_{PT}^{offset}$ (MPa) | $\sigma_{PT}^{ave}$ (MPa) | $\varepsilon_{PT}^{ave}$ | $W_{PT}^{ave}$ (×10⁶ J/m³) |
|---|---|---|---|---|---|
| Sample 1 | 752 | 1321 | 1036.5 | 0.055 | 56.6 |
| Sample 2 | 700 | 1286 | 993 | 0.052 | 51.5 |
| Sample 3 | 673 | 1269 | 971 | 0.0575 | 55.8 |
| Sample 4 | 690 | 1252 | 971 | 0.0546 | 53.0 |
| DFT | 852 | 1711 | 1281 | 0.079 | 101.2 |

**Supplementary Table 1: Calculation of actuation work per-unit-volume ( $W_{PT}^{ave}$ ).** Here, we calculate $W_{PT}^{ave}$ by $\sigma_{PT}^{ave} \times \varepsilon_{PT}^{ave}$, where $\sigma_{PT}^{ave}$ and $\varepsilon_{PT}^{ave}$ are the average transformation stress and strain, respectively.

**Supplementary Note 1: 100 cyclic compression test**

We performed the 100 cyclic test with 20 nm/sec of displacement rate and 6 mN of maximum load and 0.6 mN of minimum load, and the time-load curve in the inset clearly shows that our micropillar experiences all three stages (I, II and III) for each cycle (Supplementary Fig. 1). So, the forward and backward phase transformation should occur every cycle[3]. SEM images before and after the cyclic test confirms no height change. Thus, $CaFe_2As_2$ does not shows any residual deformation after 100 cycles, which implies the fatigue resistance better than that shape memory ceramic micropillars that typically fails before 5~30 cycles of forward-backward transition. As discussed in the manuscript, cT transition occurs by simple bond adhesion between As layers, which would not introduce a significant stress concentration in the course of phase transformation. Also, we expect a superior fatigue resistance.

**Supplementary Note 2: Stress-Strain Calculations**

In order to obtain the precise strain value in the micropillar, it is important to measure the displacement of the base under the micropillar because the total measured displacement ($u_{tot}$) includes both the displacement of micropillar ($u_{pillar}$) and the base ($u_{base}$) (Supplementary Fig. 2),

$$u_{pillar} = u_{tot} - u_{base}. \tag{1}$$

The displacement of the base can be calculated by using Sneddon flat punch solution by assuming that cylindrical flat punch of CaFe$_2$As$_2$ indents the elastic half space of CaFe$_2$As$_2$[1]. The contact stiffness ($k_{Sneddon}$) from Sneddon solution gives

$$k_{Sneddon} = \frac{2E}{(1-v^2)}\sqrt{\frac{A_{contact}}{\pi}}, \tag{2}$$

where $E$ is the Young's modulus of CaFe$_2$As$_2$, $v$ is the Poisson's ratio of CaFe$_2$As$_2$, and $A_{contact}$ is the contact area between micropillar and the base. The displacement of the base can be given by

$$u_{base} = \frac{P}{k_{Sneddon}} = \frac{P}{\frac{2E}{(1-v^2)}\sqrt{\frac{A_{contact}}{\pi}}} = \frac{P(1-v^2)}{2E}\sqrt{\frac{\pi}{A_{contact}}}, \tag{3}$$

where $P$ is the applied force. Then, the engineering strain can be calculated by

$$\varepsilon_{eng} = \frac{u_{pillar}}{L_o} = \frac{u_{tot} - u_{base}}{L_o} = \frac{1}{L_o}\left[u_{tot} - \frac{P(1-v^2)}{2E}\sqrt{\frac{\pi}{A_{contact}}}\right], \tag{4}$$

where $L_o$ is the initial height of micropillar.

Here, please note that caution is needed when we use Young's modulus because it is not correct if we use simply the Young's modulus of tetragonal phase only. We must consider the contributions of elastic displacement of the tetragonal phase, elastic displacement of the collapsed tetragonal phase, and the phase transformation displacement to the total displacement of the base. Thus, it is necessary to measure the effective Young's modulus that includes all these contributions under the condition of indentation, and it is possible to obtain it by performing nanoindentation on $CaFe_2As_2$ along c-axis. Poisson's ratio of intermetallic compound ranges typically from 0.2~0.3. Here we assumed $\nu = 0.2$. Note that the contact stiffness of the Sneddon solution is not sensitive to Poisson's ratio, so our Poisson's ratio ($\nu = 0.2$) is a reasonable approximation. With all of this in mind, we were able to obtain the effective Young's modulus by using the Oliver-Pharr method[2]. The nanoindentation data is shown in Supplementary Fig. 3(a), and the average Young's modulus is about 48 GPa. Our DFT simulation shows that Young's modulus of orthorhombic phase, which would be similar with tetragonal phase, is about 33 GPa, and that of collapsed tetragonal phase is about 103 GPa. It makes sense that our result (48 GPa) is in between these values because the tetragonal and collapsed tetragonal phases co-exist in the base during the nanoindentation. Then, the stress-strain curve can be corrected by using Supplementary Equation S4 (Supplementary Fig. 3(b)). All stress-strain curves in this work were corrected in this manner (Supplementary Fig. 3(c)).

**Supplementary Note 3: DFT Calculations**

**Unit Cell Compression**

In order to understand the uniaxial deformation of orthorhombic $CaFe_2As_2$, we used DFT to simulate the uniform straining of a single unit cell. It is important to note that at room temperature, the conditions in which the experiments were conducted, the phase transitions observed occur between the paramagnetic tetragonal phase and the non-magnetic collapsed tetragonal phase. All of the DFT calculations, which are done at 0K, involve the transition between the antiferromagnetic orthorhombic phase and the collapsed tetragonal phase. Here, we use the orthorhombic phase as a surrogate for the high temperature paramagnetic tetragonal phase since these two phases have similar lattice constants, with c being the most important, and should have similar bonding with the largest difference being the disordering of the magnetic moments, which cannot be simulated using DFT with current computer resources. The use of the orthorhombic phase as a surrogate for the tetragonal phase will include some error, and thus all the DFT simulation results should be interpreted as an approximation rather than true one-to-one match with experiments.

Our uniaxial deformation studies differ from previous ones in that instead of conducting a simulation where the uniaxial stress was held at fixed value and the lateral stresses were set to zero, these simulations are conducted with mixed boundary conditions. We prescribe a uniaxial strain along the $CaFe_2As_2$ c-axis and the stress is relaxed along the a- and b-axes. This allows us to specifically probe the behavior of $CaFe_2As_2$ as a function of strain and removes some of the issues that can arise with prescribe stress paths during mechanical deformation. However, we note that VASP is incapable of performing such calculations and a script was written to specifically adjust the lateral dimensions to reduce the lateral stresses to zero. The resulting stress – engineering strain curve is shown in Supplementary Fig. 4.

**Composite Stress-Strain Curve**

In order to investigate the transition from orthorhombic (O) to collapsed tetragonal (cT) phases in DFT (tetragonal to collapsed tetragonal in experiments) we have attempted to model the response of the CaFe$_2$As$_2$ pillar assuming that it can be comprised of two phases: O and cT. We assume that at a given macroscopic strain, $\varepsilon$, the system will assume the composition of the O and cT phases that minimizes the free energy under the assumption of mechanical equilibrium. We choose the a free energy that is analogous to the enthalpy under uniaxial stress, rather than internal energy, to model phase evolution under constant load, which is consistent with the load-controlled nature of the nanoindentor. Note that this free energy, which we call F, would correspond to the enthalpy if we considered the hydrostatic stress instead of uniaxial stress. We write this free energy (per unit CaFe$_2$As$_2$) as a function of the uniaxial stress and strain as well as the internal energy, obtained from DFT, as a function of the strain

$$F = \varphi_O E_O(\varepsilon_O) + \varphi_{cT} E_{cT}(\varepsilon_{cT}) - \sigma \varepsilon \Omega_0, \tag{5}$$

where $\varphi_i$ is the phase fraction (i.e. the fraction of CaFe$_2$As$_2$ that is in phase "i"), $E_i$ is the internal energy-per-unit phase (i.e. the energy per number of CaFe$_2$As$_2$), $\varepsilon_i$ is the strain in the "i" phase and $\Omega_0$ is the reference atomic volume of CaFe$_2$As$_2$. Here, we use the volume of the orthorhombic CaFe$_2$As$_2$ phase from our DFT simulations, which has lattice constants

a=5.60 Å, b=5.49 Å, and c=11.48 Å, resulting in $\Omega_0 = 353$ Å$^3$. The energies of the two phases have been computed as a function of uniaxial strain as shown in Figure S8. Mechanical equilibrium of the two phases requires that the constraint:

$$\sigma = \frac{\partial e_O}{\partial \varepsilon_O} = \frac{\partial e_{cT}}{\partial \varepsilon_{cT}}, \tag{6}$$

is satisfied, where $e_i$ is the strain energy per unit volume. Minimization of this free energy results in the appropriate volume fractions of the O and cT phases as well as the stress in the system under constant load. The strains in the individual phases can be determined using this phase fraction as:

$$\varepsilon = \varphi_O \varepsilon_O + \varphi_{cT} \varepsilon_{cT}. \tag{7}$$

In this definition, the strain in the cT phase, then, is the change in length per unit length in the cT phase relative to the O phase. This makes the mathematical implemnation easiest since both phases have the same reference length.

The minimization of the total free energy requires smooth functions while the DFT data, as shown in Supplementary Figs. 4 and 5, are discrete. To remedy this, we use a spline interpolant of the data to make suitable for numerical optimization. This is done by first selecting a fixed stress value and an estimate of the volume fractions of the phases which can be used, in conjunction with the Supplementary Equations (7) and (8) the numerical values for Supplementary Equation (5) can be determined. The minimization itself was carried out using the golden ratio search to find the minimum of Supplementary Equation (5) with the bounds that the phase fractions between 0 and 1. The evolution of the two phase fractions as a function of the applied strain obtained from this minimization are shown in Supplementary Fig. 6 demonstrating a non-linear relationship between phase fractions and the applied strain. The evolution of the strains are plotted in Supplementary Fig. 7 clearly showing that the strains add up to the total strain.

The full solution of these equations results in a stress-strain curve shown in Fig. 2(c) of the manuscript. It is important to note a couple of points. First, the analysis uses DFT values of the O and cT phases at zero Kelvin, and has not been corrected for higher temperatures observed in the experiments. In addition, the DFT calculations have a stable cT structure at 0K, however in the experiments the cT structure spontaneously transforms back to the O phase as stress is released as the cT phase is not meta-stable at room temperature. Our composite model predicts the same thing because it assumes that the transition occurs to minimize the total free energy and ignores energy barriers. We expect some differences between theoretical models and experiments at lower temperatures.

**Supplementary Note 4: Actuation work per-unit-volume**

Even though the actuation testing has not been performed, it is still possible to calculate the actuation work per-unit-volume from a stress-strain curve as suggested in Lai, A., et al. *Science* **27,** 1505-1508 (2013)[4], and Huber, J. E., et al. *Proc. R. Soc. London Ser. A* **453,** 2185–2205 (1997)[5] (Note that Fig. 3 came from these two references.). They calculated the actuation work of shape memory alloys and ceramics simply by (transformation stress) × (transformation strain) because the actuation work per-unit-volume can be defined as the rectangular area below the stress-strain curve within the range of transformation strain. However, in the case of $CaFe_2As_2$, because the transformation stress increases in stage II, we can estimate the actuation work per-unit-volume as the area of trapezoid. This can be simply done by (average transformation stress) × (transformation strain). Here, the average transformation stress ($\sigma_{PT}^{ave}$) is the average of onset ($\sigma_{PT}^{onset}$) and offset ($\sigma_{PT}^{offset}$) transformation stress.

The onset and the offset of transformation can be defined by using the line-intercept methods. Each blue line in Supplementary Fig. 8(a) can be obtained by taking the linear regression on carefully selected linear-like region. Once we get the onset and offset stress/strain, we can approximate the area of stage II by (average transformation stress) × (transformation strain). The data points in Fig. 3 were calculated by this method from four experimental stress-strain curves and one DFT stress-strain curve in Supplementary Fig. 8(b). The calculated the actuation work per-unit-volume data are available in Supplementary Table 1.

**Supplementary Note 5: Preliminary results on LaRu$_2$P$_2$**

We recently tested LaRu$_2$P$_2$, which is another ThCr$_2$Si$_2$-type intermetallic compound (Supplementary Fig. 9(a)). This material also undergoes collapsed tetragonal phase transformation, and exhibit superelasticity. One of the notable mechanical properties is the ultra-high yield strength (~7 GPa) (Supplementary Fig. 9(b)). Thus, regardless of superelastic strain smaller than that of CaFe$_2$As$_2$, the actuation work is still excellent due to the ultra-high yield strength. Note that there is a region of decreasing contact stiffness (Supplementary Fig. 9(c)). Note that the decrease in contact stiffness is an another evidence of structural collapse because collapsed tetragonal phase is much stiffer than tetragonal phase. Our preliminary results show that superelasticity associated with collapsed tetragonal phase transition found in CaFe$_2$As$_2$ is only one manifestation of a wider class of such transitions found in significant number of ThCr$_2$Si$_2$-type intermetallic compounds. This is very important to search for industrial applicable (non-toxic) superelastic intermetallic compounds.

**Supplementary References**